\documentclass[
conference, 
letterpaper,
twocolumn,
10pt,
]
{IEEEtran}

\usepackage{cite}
\usepackage{./mymathmacros}
\usepackage{subfig}
\usepackage[nolist]{acronym}
\usepackage{hyperref}
\usepackage{cleveref}
\usepackage{tikz}
\usepackage{pgfplots}
\usetikzlibrary{decorations.markings}
\usetikzlibrary{shapes}
\usepackage{amsmath}
\usepackage{siunitx}
\usepackage{algorithmicx}
\usepackage{algorithm}
\usepackage{algpseudocode}
\algnewcommand\algorithmicinput{\textbf{Input:}}
\algnewcommand\Input{\item[\algorithmicinput]}
\algnewcommand\algorithmicoutput{\textbf{Output:}}
\algnewcommand\Output{\item[\algorithmicoutput]}

\usepackage{csquotes}
\usepackage{amssymb}
\usepackage{multirow}
\usepackage[font=footnotesize, labelfont=bf]{caption}
\usepackage{booktabs}

\usetikzlibrary{backgrounds}
\usetikzlibrary{fit}

\crefname{equation}{}{}
\crefname{figure}{Fig.}{Figs.}
\crefname{section}{Section}{Sections}
\crefname{algorithm}{Algorithm}{Algorithms}

\IEEEoverridecommandlockouts 

\title{Quantized Message Passing for LDPC Codes\thanks{Funding by WWTF Grant ICT12-054.}}

\author{
    \IEEEauthorblockN{
       Michael Meidlinger\IEEEauthorrefmark{1},
       Alexios Balatsoukas-Stimming\IEEEauthorrefmark{2},  
       Andreas Burg\IEEEauthorrefmark{2}, and
       Gerald Matz\IEEEauthorrefmark{1}
    }
    \\
    \begin{tabular}[t]{c@{\extracolsep{3cm}}c} 
      \IEEEauthorrefmark{1}Vienna University of Technology, Austria  & \IEEEauthorrefmark{2}EPFL, Switzerland \\
        \small Email: \{\href{mailto:mmeidlin@nt.tuwien.ac.at}{mmeidlin},
                        \href{mailto:gmatz@nt.tuwien.ac.at}{gmatz}\}@nt.tuwien.ac.at 
        &
        \small Email: \{\href{mailto:alexios.balatsoukas@epfl.ch}{alexios.balatsoukas},
                       \href{mailto:andreas.burg@epfl.ch}{andreas.burg}\}@epfl.ch
                \\ 
   \end{tabular}
}

\begin{document}

\begin{acronym}
   \acro{MP}{message passing}
   \acro{PMF}{probability mass function}
   \acro{PDF}{probability density function}
   \acro{BICM}{bit-interleaved coded modulation}
   \acro{BI-AWGN}{binary input additive white Gaussian noise}
   \acro{BP}{belief propagation}
   \acro{DE}{density evolution}
   \acro{MS}{min-sum}
   \acro{SP}{sum-product}
   \acro{SNR}{signal to noise ratio}
   \acro{BSC}{binary symmetric channel}
   \acro{VN}{variable node}
   \acro{CN}{check node}
   \acro{BER}{bit error rate}
   \acro{BEP}{bit error probability}
   \acro{FER}{frame error rate}
   \acro{FAID}{finite alphabet iterative decoder}
   \acro{IBM}{information bottleneck method}
   \acro{DMC}{discrete memoryless channel}
   \acro{LLR}{log-likelihood ratio}
   \acro{LDPC}{low-density parity-check}
   \acro{LUT}{look-up table}
   \acro{MAP}{maximum a-posteriori probability}
   \acro{MI}{mutual information}
\end{acronym}

\begin{acronym}
  \acro{LDPC}{Low-density parity-check} 
  \acro{SNR}{signal-to-noise ratio} 
\end{acronym}

\maketitle
\begin{abstract}
   We propose a quantized decoding algorithm for low-density parity-check codes where the \acl{VN} update rule of the standard \acl{MS} algorithm is replaced with a \acl{LUT} (LUT) that is designed using an information-theoretic criterion. We show that even with message resolutions as low as 3 bits, the proposed algorithm can achieve better error rates than a floating-point \acl{MS} decoder. Moreover, we study in detail the effect of different decoder design parameters, like the design SNR and the LUT tree structure on the performance of our decoder, and we propose some complexity reduction techniques, such as LUT re-use and message alphabet downsizing.
\end{abstract}

\section{Introduction}\label{sec:introduction}

\noindent\ac{LDPC} codes 
have excellent error rate 
performance and can be efficiently decoded 
using \ac{MP} schemes like the \ac{SP} and the \ac{MS} algorithms, both of which
involve real-valued (infinite precision) messages. By contrast, practical implementations
require finite-precision message representations,
i.e., the decoder messages have to be quantized and represented using 
typically $4$ to $7$ bits per message.
Lower message resolutions tend to deteriorate the error rate performance of the code severely,
especially in the error floor regime at high \ac{SNR} 
\cite{zhang2014a}.
Recent work on quantized \ac{MP} decoders
\cite{planjery2013a,balatsoukas-stimming2015a,kurkoski2008a}
has shown that a significant reduction of the message resolution is possible if the decoding algorithm is 
explicitly tailored to finite message alphabets. 

In this paper, we present a novel \enquote{min-LUT}  algorithm that replaces the \ac{VN} update of the \ac{MS} algorithm with a
\ac{LUT} designed to maximize the local information flow through the code's Tanner Graph~\cite{kurkoski2014a}. In our previous work \cite{balatsoukas-stimming2015a}, we 
have shown that an actual implementation of the min-LUT decoder 
reduced the hardware complexity 
and increases the decoder throughput relative to a conventional adder-based MS implementation. 
This paper complements \cite{balatsoukas-stimming2015a} by providing 
an in-depth discussion of the algorithm and decoder design. 
Specifically, we discuss
in detail the symmetry requirements and the information-theoretic construction of the \acp{LUT} for the message updates.
Furthermore, we examine the effects of 
design \ac{SNR} and LUT tree structure on the error performance and develop additional complexity reduction techniques such as LUT reuse and alphabet downsizing. We demonstrate that LUT reuse is attractive for implementations and can even improve error rate performance. 
Finally, we show simulation results illustrating the design and performance trade-offs.

\section{\acs{LDPC} Codes and MP Decoding}\label{sec:ldpc}

A $(d_v,d_c)$-regular \ac{LDPC} code with parity-check matrix $\mat H \in \{0,1\}^{M\times N}$ can be represented by a Tanner graph consisting of $N$ \acp{VN} and $M$ \acp{CN}. Every \ac{CN} is connected to $d_c$ \acp{VN} and every \ac{VN} is connected to $d_v$ \acp{CN}, where a connection between two nodes is indicated by a non-zero entry in the parity check matrix.

\ac{LDPC} codes are traditionally decoded using \ac{MP} algorithms, where messages are exchanged between 
\acp{VN} and \acp{CN} over the course of several decoding iterations. Let $\mathcal{M}_i$ denote the message alphabet at iteration $i$.
At each iteration the messages from 
\acs{VN} $n$ to \acs{CN} $m$ are computed using the mapping 
$\Phi_v^{(i)}: \set L \times \set M^{d_v-1}_{i-1}\rightarrow \set M_i$, which is defined as\footnote{To simplify notation, we 
suppress the iteration index $i$ in the messages.}
\begin{align} \label{eqn:vnupdateGeneric}
        \mu_{n\rightarrow m} = \Phi_v^{(i)} \big(L_n, \vec{\bar\mu}_{{\mathcal N}(n)\setminus m \rightarrow n} \big),
\end{align}
where $\set N(n)$ is the set of neighbours of node $n$ in the Tanner graph, 
$\vec{\bar\mu}_{{\mathcal N}(n)\setminus m \rightarrow n} \in \set M^{d_v-1}_{i-1}$ 
is a vector containing the incoming messages from all neighboring \acp{CN} except $m$, and $L_n \in \set L$ denotes the channel \ac{LLR}  
at \ac{VN} $n$.
Similarly, the \acs{CN}-to-\acs{VN} messages are computed via the mapping $\Phi_c^{(i)}: \set M^{d_c-1}_i\rightarrow \set M_i$
defined as
\begin{equation} \label{eqn:cnupdateGeneric}
    \bar\mu_{m\rightarrow n} = \Phi_c^{(i)} \big(\vec{\mu}_{    {\mathcal N}(m)\setminus n \rightarrow m} \big).
\end{equation}
\Cref{fig:message_updates} illustrates the message updates in the Tanner graph. 
The decision for a code bit $c_n$ is computed 
with a mapping $\Phi_d: \set L \times \set M^{d_v}_I\rightarrow \{0,1\}$ 
based on the incoming check node messages and the channel \ac{LLR} 
according to
\begin{equation}\label{eqn:decupdate}
   \hat c_n = \Phi_d (L_n, \vec{\bar \mu}_{\set N(n)\rightarrow n}).
\end{equation}
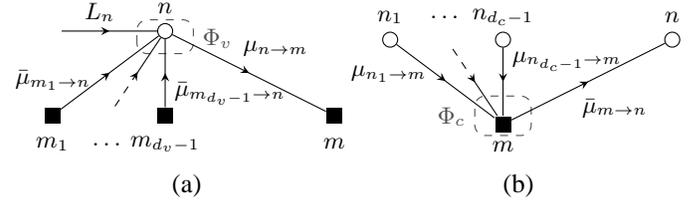
\begin{figure}%
   \centering
   \begin{tabular}{cc}
      \hspace*{-.3cm} 
\begin{tikzpicture}[
   >=stealth,
   scale = .75,
   decoration={
    markings,
    mark=at position 0.5 with {\arrow{>}}}
    ] 

\tikzstyle{every node}=[font=\small]
\tikzstyle{cnode}=[rectangle, inner sep = 3pt, fill=black]
\tikzstyle{vnode}=[draw=black, circle, inner sep =2pt]
   \draw (0   , 0) node[cnode] (m1) {};
   \draw (0   , -.5) node{$m_1$};

   \node (mphantom)  at (1   , 0) {};
   \draw (1   , -.5) node{$\dots$};

   \draw (2   , 0) node[cnode] (md) {};
   \draw (2   , -.5) node{$m_{d_v-1}$};

   \draw (5   , 0) node[cnode] (m) {};
   \draw (5   , -.5) node{$m$};

   \draw (2   , 1.5)   node[vnode] (n) {};
   \draw (2   , 1.9)  node{$n$};
   \draw[dashed, rounded corners=4pt, black!70]   (1.5   , 1.1) rectangle (2.5, 1.7) node[below right] {$\Phi_v$};

   \node (llr)  at (0   , 1.5) {};

   \draw[postaction={decorate}] (m1) -- (n) node[pos =.4, left]{$\bar\mu_{m_1\rightarrow n}$};
   \draw[postaction={decorate}, dash pattern=on 3pt off 3pt on 3pt off 3pt on 500pt] (mphantom) -- (n);
   \draw[postaction={decorate}] (md) -- (n) node[pos =.2, right]{$\bar\mu_{m_{d_v-1}\rightarrow n}$};
   \draw[postaction={decorate}] (n) -- (m) node[pos =.4, above right]{$\mu_{n\rightarrow m}$};

   \draw[postaction={decorate}] (llr) -- (n) node[pos =.4, above]{$L_n$};

\end{tikzpicture}
 & \hspace*{-.7cm} 
\begin{tikzpicture}[
   >=stealth,
   scale = .75,
   decoration={
    markings,
    mark=at position 0.5 with {\arrow{>}}}
    ] 

\tikzstyle{every node}=[font=\small]
\tikzstyle{cnode}=[rectangle, inner sep = 3pt, fill=black]
\tikzstyle{vnode}=[draw=black, circle, inner sep =2pt]
   \draw (0   , 1.5) node[vnode] (n1) {};
   \draw (0   , 1.9) node{$n_1$};

   \node (nphantom)  at (1   , 1.5) {};
   \draw (1   , 1.9) node{$\dots$};

   \draw (2   , 1.5) node[vnode] (nd) {};
   \draw (2   , 1.9) node{$n_{d_c-1}$};

   \draw (5   , 1.5) node[vnode] (n) {};
   \draw (5   , 1.9) node{$n$};

   \draw (2   , 0)   node[cnode] (m) {};
   \draw (2   , -.4)  node{$m$};
   \draw[dashed, rounded corners=4pt, black!70]   (2.5   , .5 ) rectangle (1.5, -.2) node[above left] {$\Phi_c$};

   \draw[postaction={decorate}] (n1) -- (m) node[pos =.4, left]{$\mu_{n_1\rightarrow m}$};
   \draw[postaction={decorate}, dash pattern=on 3pt off 3pt on 3pt off 3pt on 500pt] (nphantom) -- (m);
   \draw[postaction={decorate}] (nd) -- (m) node[pos =.2, right]{$\mu_{n_{d_c-1}\rightarrow m}$};
   \draw[postaction={decorate}] (m) -- (n) node[pos =.4, below right]{$\bar\mu_{m\rightarrow n}$};

\end{tikzpicture}
 \\
      (a) & \hspace*{-.5cm}(b) \\
   \end{tabular}
      \caption{
      \acs{VN} update 
      (a)
      and \acs{CN} update 
      (b)
      for $\set N (n) =\{m,m_1,\dots, m_{d_{v}-1}\}$ and $\set N (m) =\{n,n_1,\dots, n_{d_{c}-1}\}$ 
      }%
      \label{fig:message_updates}%
\end{figure}
For the \ac{MS} algorithm, the mappings read
\begin{align}
   \label{eqn:vnupdateMS}
   &\Phi_v^{\mathrm{MS}}  (L, \vec{\bar\mu}  \big)= L + \sum_i \bar\mu_i, \\ 
   \label{eqn:cnupdateMS}
   & \Phi_c^{\mathrm{MS}} (\vec \mu \big)
        = \sign(\vec \mu) \min |\vec \mu|, 
\end{align}
with $\min |\vec \mu|$ denoting the minimum of the absolute values of the vector elements and $\sign( \vec \mu )= \prod_j \sign(\mu_j)$. The \ac{MS} mappings remain unchanged for all iterations 
and all message alphabets are taken to be the reals,
$\set M_i = \set M = \set L = \mathbb R$.
The bit decision $\Phi_d$ is based on the sign of the a-posteriori LLRs,
\begin{equation} \label{eqn:minsumDec}
   \Phi_d^{\mathrm{MS}}(L,\vec{\bar\mu}) =  \frac{1}{2}\bigg(1-\sign\!\bigg (L + \sum_i \bar\mu_i \bigg)\bigg).
\end{equation}

\section{The Min-LUT Algorithm}\label{sec:decoder_design}

\subsection{Basic Idea}\label{subsec:minLUTbasic}

Since floating-point arithmetic is not feasible for practical hardware implementations,
the real-valued messages of the MS algorithm are usually discretized using a small number of uniformly spaced quantization levels.
Together with the well-established two's complement and sign-magnitude binary encoding,  
the uniform quantization leads to efficient arithmetic circuits but leads to degraded error-rate performance. 

Recently, efforts have been made to design decoders that explicitly account for finite message and \ac{LLR} 
alphabets~\cite{planjery2013a,kurkoski2008a}. Instead of arithmetic computations such as \cref{eqn:vnupdateMS} and 
\cref{eqn:cnupdateMS}, the update rules for these decoders are implemented as \acfp{LUT}. 
There are numerous approaches to the design of such LUTs. In the following, we present an algorithm that 
combines the conventional \ac{MS} algorithm and the purely \ac{LUT}-based approach of \cite{kurkoski2008a}.
In this  min-LUT algorithm, the VN updates are realized as \acp{LUT}, whereas the CN updates follow 
\cref{eqn:cnupdateMS}. This is motivated by the following observations:
\begin{itemize}
   \item The CN degree is larger than the VN degree, especially for high code rates. Consequently, without further
      simplifications, the \ac{CN}
      \acp{LUT} are far more complex than \ac{VN} \acp{LUT} as the size of the \acp{LUT} grows exponentially in the number
      of inputs.
   \item For the \ac{MS} algorithm, the 
   VN update \cref{eqn:vnupdateMS} typically increases the dynamic range of the messages 
      whereas the CN update \cref{eqn:cnupdateMS} preserves the dynamic range. 
      Replacing the VN update \cref{eqn:vnupdateMS} with a LUT eliminates the
      need for a message representation that can be interpreted as a numeric value. As will be explained in \cref{subsec:channel_symmetry}, the outputs of the \ac{LUT}-based \acp{VN} can be sorted in such a way that the CN update \cref{eqn:vnupdateMS} can be performed based on the LUT output labels. 
\end{itemize}

The LUT design for the VN updates  is based on \cite{kurkoski2008a} and follows a \ac{DE} approach.
Given the CN-to-VN message distributions 
of the previous iterations, one can design the VN LUTs for each iteration in a way that maximizes the mutual information 
between the VN output messages and the codeword bit corresponding to the VN in question. 

In order to initialize the \ac{DE} procedure, we first characterize the  LLR distribution at the decoder input in \cref{subsec:channel_symmetry}. Furthermore, \cref{subsec:channel_symmetry} discusses the relevance of symmetry conditions for the min-LUT algorithm.
After 
these prerequisites, we present the actual evolution of message \acp{PMF} in \cref{subsec:IT}.

\subsection{Channel Model and Symmetry Conditions}\label{subsec:channel_symmetry}

Throughout this  paper, we focus on a  \ac{BI-AWGN} channel $p_{\rv y|\rv x}(y|x)$ with noise variance $\sigma^2$ followed by a  quantizer  $Q_L: \mathbb R \rightarrow \set L$. 
The quantizer uses an even number of levels $|\set L|$ and the quantization regions are symmetric about the origin.
The quantized LLRs  are derived from the output of the \ac{BI-AWGN} channel via
$L = Q_L \left( -\frac{2 y}{\sigma^2} \right)$,
inducing a symmetric pmf $p_{\rv L | \rv x}(L|x)$
that can in turn be used to define the reproducer values of the quantized LLRs as
\begin{equation}\label{eqn:llrVal}
   L \defas \log \frac{p_{\rv L | \rv x}(L|0)}{p_{\rv L | \rv x}(L|1)},
\end{equation}
hence $p_{\rv L | \rv x}(-L|0) = p_{\rv L | \rv x}(L|1)$
Similarly, we can assign  reproducer values 
\begin{equation}\label{eqn:msgVal}
   \mu \defas \log \frac{p^{(i)}_{\rv m | \rv x}(\mu |0)}{p^{(i)}_{\rv m | \rv x}(\mu |1)}
\end{equation}
to the output message labels of the VN LUTs at iteration $i$. We again assume 
that the number $|\set M|$ of messages is even. 
When the reproducer values are in an ascending order,
\begin{equation}\label{eqn:ordering}
   L_1 < L_2 < \dots < L_{|\set{L}|}, \qquad \mu_1 < \mu_2 < \dots < \mu_{|\set{M}|},
\end{equation}
the identities
\begin{equation}\label{eqn:llr_symmetry_disc}
   L_k  \equiv -L_{|\set L|-k+1},  \qquad  \mu_j  \equiv -\mu_{|\set M|-j+1},
\end{equation}
follow from the symmetry of $p_{\rv L | \rv x}(L|x)$ and the \ac{MP} algorithm (cf. \cite{richardson2001a}, Definition 1) 
and associate each label $k \in\{1,\dots,|\set L|\}$ and $j \in\{1,\dots,|\set M|\}$ with a sign.
Based on this association and the ordering \cref{eqn:ordering}, the \ac{MS} \ac{CN} update \cref{eqn:cnupdateMS}
can be performed directly on the message labels; the reproducer values \cref{eqn:llrVal,eqn:msgVal} are not needed for
decoding. However, \cref{eqn:msgVal} bears an interesting interpretation: As the messages become more informative over the course of iterations ---implying more concentrated densities  $p^{(i)}_{\rv m | \rv x}(\mu |x)$--- the reproducer values
grow in magnitude. Using different LUTs for different iterations is thus similar to using different message representations
for different iterations, an approach which has already been used successfully in \cite{zhang2014a}.

The symmetry of the \ac{MP} algorithm discussed above is guaranteed, if the \ac{VN} LUT at any iteration $i$ 
satisfies
\begin{equation}\label{eqn:vnSymmetry}
   \!\!\!\Phi_v^{(i)}(-L,-\bar\mu_1, \dots, -\bar\mu_{d_v-1})\! = -\Phi^{(i)}_v(L,\bar\mu_1, \dots, \bar\mu_{d_v-1}). 
\end{equation}
This identity can be reformulated based on \cref{eqn:llr_symmetry_disc} as a symmetry relation involving only labels.

Whereas our decoder design is exemplified for the BI-AWGN channel, it applies to any symmetric binary input channel followed by a symmetric quantizer. As an example, the channels characterized in \cite{alvarado2009b} could be used to design decoders for \ac{BICM} systems.

\subsection{Density Evolution and \ac{LUT} Design} \label{subsec:IT}

In this section, we show how the message \acp{PMF} evolve over the course of iterations.
We first describe how the distribution of the CN-to-VN messages can be computed based on the distribution of the incoming 
VN-to-CN messages. If the Tanner graph is cycle-free, then the individual input messages of a \ac{CN} at iteration
$i$ are iid conditioned on the transmitted bit $\rv x$, and their distribution is denoted by $p^{(i)}_{\rv m | \rv x} (\mu | x)$. 
The joint distribution of the $(d_c-1)$ incident messages 
conditioned on the transmitted bit value corresponding to the recipient \ac{VN} 
(cf. Fig.~\ref{fig:message_updates}) reads
\begin{equation}\label{eqn:CNprodChannel}
   p^{(i)}_{\vec{{\rv m}} | \rv x} (\vec\mu | x) =
     \left(\frac{1}{2}\right)^{dc-2} 
   \sum_{\vec x:\, \bigoplus \vec x = x}
   \prod_{j=1}^{d_c-1}
      p^{(i)}_{\rv m | \rv x} (\mu_j | x_j),
\end{equation}
where $\bigoplus \vec x$ denotes the modulo-$2$ sum of the components of $\vec x$.
Using the update rule \cref{eqn:cnupdateMS}, the distribution of the outgoing
\acs{CN}-to-\acs{VN} message is then given by
\begin{equation} \label{eqn:CNmsgDist}
   p^{(i)}_{\overline{\rv m} | \rv x}(\bar\mu | x) =
   \sum_{\vec \mu:\,\sign(\vec \mu) \min |\vec \mu|=\bar\mu} 
    p^{(i)}_{\vec{{\rv m}} | \rv x} (\vec\mu | x).
\end{equation}

Let $\vec{\bar\mu} = \begin{pmatrix} \bar\mu_1, \dots, \bar\mu_{d_v-1}\end{pmatrix}$ 
denote the $(d_v-1)$ incident \ac{CN}-to-\ac{VN} messages that are involved in the update of a certain \ac{VN}. 
Then, the joint distribution of the VN input messages and the LLR is given by
\begin{equation} \label{eqn:VNprodChannel}
       p^{(i)}_{\rv L,\vec{\overline{\rv m}} | \rv x} (L,\vec{\bar\mu} | x) =
   \hspace{-17 pt}\sum_{\vec x:\, x_0=\dots= x_{d_v-1} = x}\hspace{-17pt}
   p_{\rv L| \rv x} ( L | x_0 )
   \prod_{j=1}^{d_v-1} 
   p^{(i)}_{\overline{\rv m} | \rv x}(\bar\mu_j | x_j).
\end{equation}
Given this distribution, we can construct an update rule $\Phi_{v}$ that maximizes the mutual information 
$I_i(\rv m; \rv x)$ between $\rv m$ and $\rv x$:
\begin{equation}\label{eqn:MIquantizer} 
   \Phi_{v}^{(i)} = \argmax_{\Phi} I_i(\rv m; \rv x) = 
   \argmax_{\Phi} I_i\big(\Phi( \rv L, \overline{\rvec m} ); \rv x\big).
\end{equation}
Here, the maximization is over  all deterministic mappings $\Phi$ in the form of \cref{eqn:vnupdateGeneric} that respect
the symmetry condition \cref{eqn:vnSymmetry}. 
Hence, the resulting update rule $\Phi_{v}^{(i)}$  
maximizes the local information flow between the \acp{CN} and the \acp{VN}.
An algorithm that solves \cref{eqn:MIquantizer} with complexity $O\big( |\set L|^3|\set M|^{3(d_v-1)}\big)$  was provided in~\cite{kurkoski2014a}. 
Using the update rule \cref{eqn:MIquantizer}, we can compute the
conditional distribution of the messages in the next iteration:
\begin{equation}\label{eqn:VNmsgDist}
    p^{(i+1)}_{\rv m | \rv x}(\mu | x) =
   \hspace{-15 pt}\sum_{(L,\vec{\bar\mu)}:\,\Phi_{v}^{(i)}(L,\vec{\bar\mu}) = \mu} \hspace{-15 pt}
   p^{(i)}_{\rv L,\overline{ \vec{\rv m}} | \rv x} (L,\vec{\bar\mu} | x).
\end{equation}

The noise threshold $\sigma^*$ of a $(d_v,d_c)$-regular LDPC code ensemble with at most $I$ decoding iterations is defined as
\begin{equation}
	\sigma^* = \sup\!\left\{\sigma \!\geq\! 0\!:\, I_i(\rv m; \rv x)>1\!-\!\epsilon \; \text{for some }
	 i \le I \right\}.
\end{equation}
\Cref{alg:01} summarizes the individual steps of a bisection algorithm that uses the \ac{DE} algorithm to calculate $\sigma^*$.

\begin{algorithm}
\begin{algorithmic}[1]
   \caption{Density Evolution based LUT design}\label{alg:01}
   \Input{Search interval $[\sigma_{\mathrm{min}}, \sigma_{\mathrm{max}}]$, precision $\Delta\sigma>0$, $\epsilon$, $I$}
   \While{$\sigma_{\mathrm{max}}-\sigma_{\mathrm{min}}>\Delta\sigma$}
   \State{$\sigma \gets (\sigma_{\mathrm{max}}-\sigma_{\mathrm{min}})/2$}
   \State{Get $p_{\rv L | \rv x}(L|x)$ corresponding to BI-AWGN($\sigma^2$)}
   \State{achievable $\gets$ false}
      \For{$i=1,\dots,I$}
      \State{Update CN-to-VN  distribution \cref{eqn:CNprodChannel,eqn:CNmsgDist}}
      \State{Build the product distribution \cref{eqn:VNprodChannel}}
      \State{Design LUT update $\Phi_v^{(i)}$ \cref{eqn:MIquantizer}}
      \State{Update VN-to-CN  distribution \cref{eqn:VNmsgDist} }
         \If{$ I(\rv m^{(i)}; \rv x)>1\!-\!\epsilon$}
            \State{achievable $\gets$ true}
            \State{\bf break}
         \EndIf
      \EndFor
      \If{achievable}
         \State{$\sigma_{\mathrm{min}}\gets\sigma$}
      \Else
         \State{$\sigma_{\mathrm{max}}\gets\sigma$}
      \EndIf
   \EndWhile
   \State{$\sigma^* \gets \sigma$}
   \Output{Threshold $\sigma^*$, LUT sequence $\Phi_v^{(1)},\dots,\Phi_v^{(i)}$}
\end{algorithmic}
\end{algorithm}

\section{Design and Performance Trade-offs for practical Decoders}\label{sec:design}

\cref{alg:01} is well suited to determine the asymptotic performance of the min-LUT algorithm for large block length $N$
and many decoding iterations (large $I$). 
In order to design practical min-LUT decoders with $N$ and $I$ not too large, 
we propose the following approach:

\begin{enumerate}
   \item Choose a practical number of maximum iterations $I$.
   \item Define a reuse pattern $\set I = \{i_1,\dots,i_r\} \subseteq \{ 1,\dots,I\}$.
   \item Choose a LUT tree structure, cf. \cref{sec:reuse}   
   \item Choose a design SNR $\gamma$ such that the corresponding noise level $\sigma$
      is below the threshold $\sigma^\ast$. 
   \item For the chosen $\sigma$, run the inner loop of \cref{alg:01}, (lines 3 to 14).
      However, only design a new LUT if $i \in \set I$. If $i \not\in \set I$, reuse the LUT from
      the previous iteration.
   \item Check the performance of the results by error rate simulations; possibly repeat the
      procedure with adjusted parameters.
\end{enumerate}
The resulting LUTs can be used to synthesize a decoder that outperforms a conventional \ac{MS} decoder in  error rate performance, throughput,
and hardware complexity \cite{balatsoukas-stimming2015a}. Since for the above procedure there are
several design parameters to be chosen, we next give an overview of the performance impact of each of the individual parameters. We support our discussion with comprehensive simulation results that illustrate the design and performance trade-offs. All simulations
have been conducted using the $(6,32)$-regular LDPC code 
(block length $N=2048$, rate $R = 13/16$)  
defined for the $10$ Gbit/s Ethernet standard~\cite{802.3an}. 

\subsection{Design SNR}\label{sec:designsnr}
The information-theoretic LUT design depends strongly on the initial LLR distribution and thus in turn on the design
SNR and the LLR quantizer. 
Our simulations indicate that even though the min-LUT decoder is designed for one particular SNR, excellent performance is maintained
over a wide range of actual noise levels, cf.~\cref{fig:fer_floor}.
Re-designing the LLR quantizer or the entire decoder for different SNRs would further improve the performance
but simultaneously would substantially increase the implementation cost. 
For this reason, we kept both the LLR quantizer and the decoder fixed over the range of simulated SNRs. 

We next discuss how the choice of the design SNR affects decoder performance. As can be seen in \cref{fig:fer_floor}, by increasing
the design SNR, we can trade off performance in the waterfall region against performance in the error floor region. The interpretation
is straight-forward: decoders that are designed for bad channels work better for bad channels and vice versa. Another interpretation
can be found in terms of decoding iterations: a lower design SNR implies that the decoder is operating closer to the \ac{DE} threshold and thus \ac{DE} convergence is much slower as compared to the case of a higher design SNR well beyond the threshold.
If, however, decoders designed for low and high $\gamma$ use the same number of iterations, the lack of convergence translates into a higher residual error for low design SNRs.

\subsection{LUT Reuse and Alphabet-Downsizing}\label{sec:reuse}
\Cref{alg:01} produces a distinct VN LUT for each iteration. While this does not affect silicon complexity for an unrolled decoder
architecture, non-unrolled decoders would need to implement multiple LUTs for the VNs. Contrary to our expectations, we found in our
simulations that
reusing LUTs for multiple iterations does not necessarily degrade the performance and can even lead to an improvement. As an example, \cref{fig:fer_reuse} shows that with a reuse pattern $\set I = \{1,5\}$ with only $r=2$ different LUTs, we can improve the error rate compared to a decoder that uses distinct LUTs for every iteration.  
An explanation for this effect is still an open issue to be explored. At this point, we can only conjecture that the effect originates from the overly optimistic message distributions of \ac{DE}, which tends to overestimate the speed of convergence for practical codes that are not cycle-free.

Another means of reducing LUT complexity is
message downsizing, i.e., reducing the size of the message set, 
\[
   |\set M_{i'}| \leq |\set M_i|\qquad \text{for}\;\, i'>i .
\] 
The idea here is that the
messages undergo a gradual hardening while being passed through the decoder before culminating
into the binary-output decision mapping \cref{eqn:decupdate}. As can be observed in 
\cref{fig:fer_reuse}, a decoder with down-sized LUTs using decaying message resolutions from 3 to 2 to 1 bits over
the range of $I=8$ iterations performs only slightly worse than a comparable min-LUT decoder 
with fixed resolution of 3 bits.
LUT reuse and LUT downsizing cannot be combined arbitrarily, i.e., 
reducing the message resolution in a certain iteration prevents reuse of the corresponding LUT.

\subsection{LUT Tree Structure}\label{sec:tree}
Since the number of input configurations for the \ac{VN} update $\Phi_v^{(i)}$ equals $|\set L||\set M_i|^{d_v-1} $, a full-fledged \ac{LUT} would be prohibitively complex for codes with high \ac{VN} degree $d_v$.
A similar problem occurs with the decision \acp{LUT} \cref{eqn:decupdate}.
To overcome this limitation, we restrict ourselves to nested update rules, e.g., for
$d_v = 6$ a possible nesting could take the form
\[
   \Phi( \bar\mu_1, \dots, \bar\mu_{5}) = 
      \Phi_1\big( \Phi_2(\bar\mu_1,\bar\mu_2,\bar\mu_3), 
                  \Phi_3(\bar\mu_4,\bar\mu_5), L\big).
\]
Obviously, any such nesting can be represented graphically by a directed tree, cf.~\cref{fig:tree_array}, tree $T_2$ for this particular example. 
Since we assume iid messages, the ordering of the arguments in the nesting is immaterial and we consider
nesting that differ only in the ordering as  equivalent. 

While the nested structure clearly reduces complexity, it is not clear a priori, 
which tree structure to prefer over another. In what follows,
we provide guidelines on how to choose the tree structure based on information-theoretic
arguments and a heuristic metric. For the moment, we do not distinguish between messages $\bar \mu$ and channel input $L$; 
the discussion of the location of $L$ within the tree is deferred to \cref{sec:llr_position}.

\subsubsection{Partial ordering}
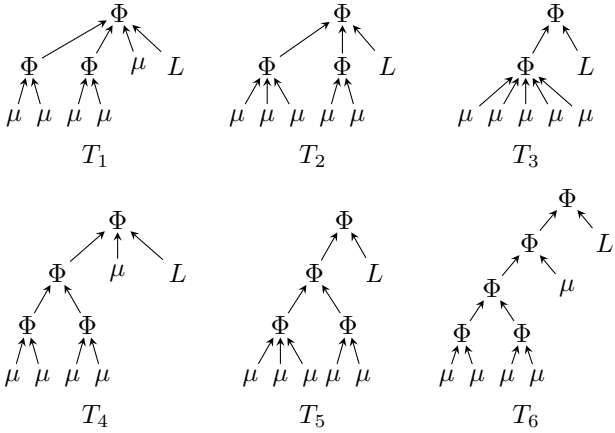
\begin{figure}[htpb]
   \centering
\tikzset{
leavenode/.style = {align=center, 
                    inner sep=2pt, 
                    text centered
                   },
imnode/.style = {align=center, 
                 inner sep=1pt, 
                 text centered, 
                }
}
\def\imstring{$\Phi$}
\def\chastring{$L$}
\def\msgstring{$\mu$}

\begin{tabular}{ccc}
\begin{tikzpicture}
[<-,
>=stealth,
level 1/.style={sibling distance=.8cm},
level 2/.style={sibling distance=.4cm},  
level distance = .7cm
] 
\node (root)[imnode] {\imstring}
   child{ node[imnode] {\imstring}
      child{ node [leavenode] {\msgstring}
      }
      child{ node [leavenode] {\msgstring}
      }
   }
   child{ node[imnode] {\imstring}
      child{ node [leavenode] {\msgstring}
      }
      child{ node [leavenode] {\msgstring}
      }
   }
   child[sibling distance=.5cm]{ node [leavenode] {\msgstring}
   }
   child[sibling distance=.5cm]{ node [leavenode] {\chastring}
   }
;
\end{tikzpicture}
 &  
\begin{tikzpicture}
[<-,
>=stealth,
level 1/.style={sibling distance=1cm},
level 2/.style={sibling distance=.4cm}, 
level distance = .7cm
] 
\node (root)[imnode] {\imstring}
   child{ node[imnode] {\imstring}
      child{ node [leavenode] {\msgstring}
      }
      child{ node [leavenode] {\msgstring}
      }
      child{ node [leavenode] {\msgstring}
      }
   }
   child{ node[imnode] {\imstring}
      child{ node [leavenode] {\msgstring}
      }
      child{ node [leavenode] {\msgstring}
      }
   }
   child[sibling distance=.6cm]{ node [leavenode] {\chastring}
   }
   ;
\end{tikzpicture}
 &  
\begin{tikzpicture}
[<-,
>=stealth,
level 1/.style={sibling distance=.8cm},
level 2/.style={sibling distance=.4cm}, 
level distance = .7cm
] 
\node (root)[imnode] {\imstring}
   child{ node[imnode] {\imstring}
      child{ node [leavenode] {\msgstring}
      }
      child{ node [leavenode] {\msgstring}
      }
      child{ node [leavenode] {\msgstring}
      }
      child{ node [leavenode] {\msgstring}
      }
      child{ node [leavenode] {\msgstring}
      }
   }
   child{ node [leavenode] {\chastring}
   }
;
\end{tikzpicture}
  \\
    $T_1$                       & $T_2$                        & $T_3$                         \\[5pt]
\begin{tikzpicture}
[<-,
>=stealth,
level 1/.style={sibling distance=.8cm},
level 2/.style={sibling distance=.8cm}, 
level 3/.style={sibling distance=.4cm}, 
level distance = .7cm
] 
\node (root)[imnode] {\imstring}
   child{ node[imnode] {\imstring}
      child{ node[imnode] {\imstring}
         child{ node [leavenode] {\msgstring}
         }
         child{ node [leavenode] {\msgstring}
         }
      }
      child{ node[imnode] {\imstring}
         child{ node [leavenode] {\msgstring}
         }
         child{ node [leavenode] {\msgstring}
         }
      }
   }
   child{ node [leavenode] {\msgstring}
   }
   child{ node [leavenode] {\chastring}
   }
;
\end{tikzpicture}
 &  
\begin{tikzpicture}
[<-,
>=stealth,
level 1/.style={sibling distance=.8cm},
level 2/.style={sibling distance=.9cm}, 
level 3/.style={sibling distance=.4cm}, 
level distance = .7cm
] 
\node (root)[imnode] {\imstring}
   child{ node[imnode] {\imstring}
      child{ node[imnode] {\imstring}
         child{ node [leavenode] {\msgstring}
         }
         child{ node [leavenode] {\msgstring}
         }
         child{ node [leavenode] {\msgstring}
         }
      }
      child{ node[imnode] {\imstring}
         child{ node [leavenode] {\msgstring}
         }
         child{ node [leavenode] {\msgstring}
         }
      }
   }
   child{ node [leavenode] {\chastring}
   }
;
\end{tikzpicture}
 &  
\begin{tikzpicture}
[<-,
>=stealth,
level 1/.style={sibling distance=1cm},
level 2/.style={sibling distance=1cm}, 
level 3/.style={sibling distance=.8cm}, 
level 4/.style={sibling distance=.4cm}, 
level distance = .6cm
] 
\node (root)[imnode] {\imstring}
   child{ node[imnode] {\imstring}
      child{ node[imnode] {\imstring}
         child{ node[imnode] {\imstring}
            child{ node [leavenode] {\msgstring}
            }
            child{ node [leavenode] {\msgstring}
            }
         }
         child{ node[imnode] {\imstring}
            child{ node [leavenode] {\msgstring}
            }
            child{ node [leavenode] {\msgstring}
            }
         }
      }
      child{ node [leavenode] {\msgstring}
      }
   }
   child{ node [leavenode] {\chastring}
   }
   ;
\end{tikzpicture}
  \\
    $T_4$                       & $T_5$                        & $T_6$                         \\
\end{tabular}
  \\
   \caption{
      Six different LUT tree structures. Note that 
      $T_1 \geq_{\set T} T_4 \geq_{\set T} T_6$,
      $T_2 \geq_{\set T} T_5$,
      $T_3 \geq_{\set T} T_5$, and
      $T_3 \geq_{\set T} T_6$. However, we cannot compare
      $T_2$ with $T_3$ or $T_5$ with $T_6$ using the relation
      $\geq_{\set T}$.
   }
   \label{fig:tree_array}
\end{figure}

Let the tree $T_1$ represent a specific nesting and let $T_2$ be a refinement\footnote{Graphically, a refinement of nesting corresponds to the placement of new nodes between parent and child nodes.} of $T_1$.
Furthermore, let $\set Q_j$ denote the set of all LUTs that respect the nesting
induced by some tree $T_j$.
By construction, any LUT in $\set Q_2$ also conforms with the nesting associated with $T_1$. Thus,
$\set Q_1 \supseteq \set Q_2$ and
\begin{equation*}
   \max_{\Phi\in \set Q_1} I_i\big(\Phi( \rv L, \overline{\rvec m}); \rv x\big) \geq
\max_{\Phi\in \set Q_2} I_i\big(\Phi( \rv L, \overline{\rvec m}); \rv x\big).
\end{equation*}
Consequently, tree refinement defines a partial ordering $\geq_{\set T}$, effectively inducing a hierarchy in terms
of maximum information flow. However, since the totality axiom is not fulfilled, 
not all tree structures can be compared in terms of the relation $\geq_{\set T}$, cf.~\cref{fig:tree_array}.

\subsubsection{A heuristic metric}
The data processing inequality states that processing can only reduce mutual information.
Therefore, for maximum information flow the paths from the input leaves to the root output should be as short as possible. 
We thus define the cumulative depth
$\lambda(T)$ of a tree $T$ as the sum of distances of all leaf nodes to the root node. 
\ac{DE} simulations confirmed that cumulative depth is useful in ranking
tree structures. \Cref{tab:01} shows how a larger $\lambda$ corresponds with a lower \ac{DE} threshold. 
However, the threshold differences are small
and our simulations have shown that all the trees presented here perform similar in terms of error rate. 
While there were small differences conforming with the ordering discussed above, they are not significant enough 
to serve as a basis for choosing the tree structure.
Rather, we recommend choosing the tree based on its silicon
complexity. Trees that are close to complete binary trees 
are preferable because they have short critical paths with low complexity LUTs and at the same time 
have small cumulative depth $\lambda$.

\begin{table}[tpb]
   \centering
   \caption{Comparison of cumulative depth and DE threshold for various
   tree structures (cf.~\cref{fig:tree_array}). 
   Here, all LUTs had a resolution of $3$ bit.
   }
   \begin{tabular}{ccccccc}
      \toprule
		$T$ & $T_1$ & $T_2$ & $T_3$ & $T_4$ & $T_5$ & $T_6$ \\
      \midrule
		$\lambda$ & 10 & 11 & 11 & 14 & 16 & 19 \\
		$\sigma^\ast$ & 0.5330  & 0.5328 & 0.5327 & 0.5313 & 0.5309  & 0.5305 \\
      \bottomrule
   \end{tabular}
      \label{tab:01}
      \vspace*{-3mm}
\end{table}

\subsubsection{Position of the channel LLR}\label{sec:llr_position}
The mutual information between the CN-to-VN messages and the coded bits is initially zero and increases over the course of iterations until at some iteration
$I_{i'}(\overline{\rv m}; \rv x) \geq I(\rv L; \rv x)$. Using a similar argument as before,
we can conclude that until iteration $i'$ the channel LLR should be placed close to the root node to ensure a large information flow. After iteration $i'$, the CN-to-VN messages tend to  carry more information than the channel LLR an thus should be placed closer to the root node. 
Our simulations show that this strategy indeed provides the best FER performance; however,
the loss as compared to the case where the channel LLR stays at the root node is only relevant
for a large number of iterations ($I>20$).

\subsection{Comparison with MS}
As can be seen in \cref{fig:fer_reuse}, the min-LUT decoder with a message resolution of 3 bits outperforms a conventional \ac{MS}
decoder using a message resolution of 4 bits by a significant margin and  even
beats a floating point \ac{MS} decoder. The gain is even larger for the case
of LUT reuse. We conclude that our min-LUT decoder is an attractive alternative
to the conventional MS decoder.

\def\mylegendfontsize{\scriptsize}

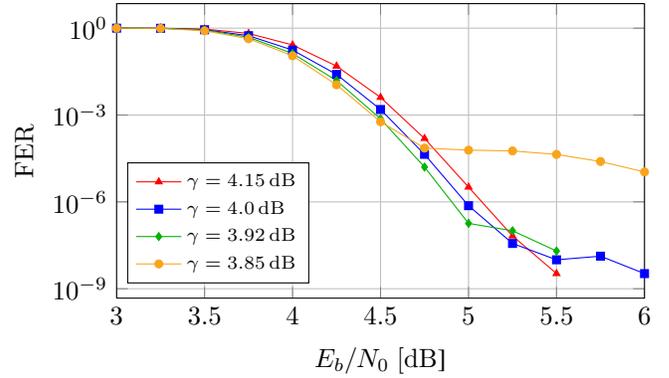
\begin{figure}[t]
   \centering
\begin{tikzpicture}
\begin{semilogyaxis}[
    width=8.6cm,
    height=5.5cm,
    xlabel={$E_b/N_0\;[\mathrm{dB}$]},
    ylabel={$\mathrm{FER}$},
    xmax = 6,
    xmin = 3,
    title={},
    mark size=1.5,   
    grid=major,          
    legend style={
        cells={anchor=west}, 
        at={(.02,.05)},        
        anchor=south west    
    }
   ]

   \addplot+[
      mark=triangle*,
      mark options={fill=red}, 
      color=red,
      ] 
      table [x index             = 0,
             y  index            = 1, 
          col sep=comma]{LUTCha3_SNR4.2_FER.csv };
   \addlegendentry{\mylegendfontsize $\gamma=4.15\,\mathrm{dB}$}

   \addplot+[
      mark=square*,
      mark options={fill=blue}, 
      color=blue,
      ] 
      table [x index             = 0,
             y  index            = 1, 
          col sep=comma]{LUTCha3_SNR4.0_FER.csv };
      \addlegendentry{\mylegendfontsize  $\gamma=4.0\,\mathrm{dB}$}

      \addplot+[
      mark=diamond*,
      mark options={fill=green!70!black}, 
      color=green!70!black,
      ] 
      table [x index             = 0,
             y  index            = 1, 
          col sep=comma]{LUTCha3_SNR3.92_FER.csv };
   \addlegendentry{\mylegendfontsize $\gamma=3.92\,\mathrm{dB}$}

      \addplot+[
      mark=*,
      mark options={fill=yellow!60!red}, 
      color=yellow!60!red,
      ] 
      table [x index             = 0,
             y  index            = 1, 
          col sep=comma]{LUTCha3_SNR3.85_FER.csv };
   \addlegendentry{\mylegendfontsize $\gamma=3.85\,\mathrm{dB}$}

\end{semilogyaxis}
\end{tikzpicture}

   \caption{\ac{FER} versus channel SNR for min-LUT decoder at different design SNRs
   ($I=8$, LUT tree $T_1$, message resolution $Q_\mu$ down-sized from 3 to 2 bit, $Q_L=3$ bit per channel LLR).}
   \label{fig:fer_floor}
\end{figure}

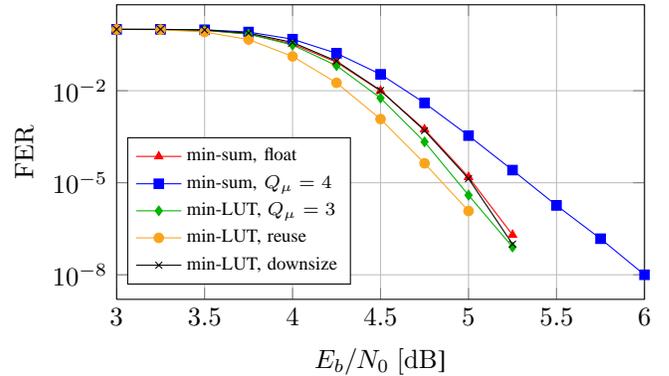
\begin{figure}[t]
\vspace*{5mm}
   \centering
\begin{tikzpicture}
\begin{semilogyaxis}[
    width=8.6cm,
    height=5.5cm,
    xlabel={$E_b/N_0\;[\mathrm{dB}$]},
    ylabel={$\mathrm{FER}$},
    xmax = 6,
    xmin = 3,
    title={},
    mark size=1.8,   
    grid=major,          
    legend style={
        cells={anchor=west}, 
        at={(.02,.05)},        
        anchor=south west    
    }
   ]

   \addplot+[
      mark=triangle*,
      mark options={fill=red}, 
      color=red,
      ] 
      table [x index             = 0,
             y  index            = 1, 
          col sep=comma]{float_FER.csv };
   \addlegendentry{\mylegendfontsize min-sum, float}

   \addplot+[
      mark=square*,
      mark options={fill=blue}, 
      color=blue,
      ] 
      table [x index             = 0,
             y  index            = 1, 
          col sep=comma]{fixedQ4_FER.csv };
      \addlegendentry{\mylegendfontsize min-sum, $Q_\mu=4$}

      \addplot+[
      mark=diamond*,
      mark options={fill=green!70!black}, 
      color=green!70!black,
      ] 
      table [x index             = 0,
             y  index            = 1, 
          col sep=comma]{LUT_noReUse_FER.csv };
   \addlegendentry{\mylegendfontsize min-LUT, $Q_\mu=3$}

   \addplot+[
      mark=*,
      mark options={fill=yellow!60!red}, 
      color=yellow!60!red,
      ] 
      table [x index             = 0,
             y  index            = 1, 
          col sep=comma]{LUT_ReUseV2_FER.csv };
          \addlegendentry{\mylegendfontsize min-LUT, reuse}
   
   \addplot+[
      mark=x,
      mark options={fill=black}, 
      color=black
      ] 
      table [x index             = 0,
             y  index            = 1, 
          col sep=comma]{LUT_downSize_FER.csv };
          \addlegendentry{\mylegendfontsize min-LUT, downsize}

\end{semilogyaxis}
\end{tikzpicture}

   \caption{Performance comparisons for different decoders
   ($I=8$, LUT tree $T_6$, $Q_L=4$, $\gamma = \SI{4.2}{dB}$).}
   \label{fig:fer_reuse}
\end{figure}

\section{Conclusion}\label{sec:conclusion}

In this paper, we presented the min-LUT algorithm for decoding LDPC codes. 
Contrary to the min-sum algorithm, the min-LUT decoder is custom-designed to
work with discrete messages of very low resolution. Hence, it constitutes
an attractive choice for practical hardware implementations. Using the
10\,Gbit/s Ethernet code, we furthermore
exemplified that the min-LUT error rate performance
can be superior to min-sum decoding 
in spite of small message resolutions.

 \small
\bibliographystyle{IEEEtran}

\end{document}